\documentclass[10pt]{article}
\usepackage{epsf,epsfig,graphics}
\newcommand{\prepr}[1] {\begin{flushright}  {\bf #1} \end{flushright}
\vskip 1.cm}
\newcommand{\titul}[1] {\begin{center}{\Large {\bf #1 } } \end{center}
\vskip 0.8cm}

\newcommand{\autor}[1] {\begin{center}  {\bf \lineskip .3cm #1  }
                        \end{center} }

\newcommand{\lugar}[1] {\begin{center}  {\normalsize \bf \it #1   }
\end{center}}
\newcounter{muni}

\begin{document}
\hbadness=10000
\pagenumbering{arabic}
\begin{titlepage}
\prepr{hep-ph/0308xxx \\IPAS-03-05 }

\titul{\bf Weak phases from topological-amplitude parametrization}

\autor{Yeo-Yie Charng$^1$\footnote{Email:
charng@phys.sinica.edu.tw} and Hsiang-nan
Li$^{1,2}$\footnote{Email: hnli@phys.sinica.edu.tw}}
\lugar{$^1$Institute of Physics, Academia Sinica, Taipei,\\
Taiwan 115, Republic of China}
\lugar{$^2$Department of Physics, National Cheng-Kung University,\\
Tainan, Taiwan 701, Republic of China}

\vskip 2.0cm
%{\bf PACS index : 13.25.Hw, 11.10.Hi, 12.38.Bx, 13.25.Ft}

\thispagestyle{empty}
%%%%%%%%%%%%%%%%%%%%%%%%%%%%%%%%%%%%%%%%%%%%%%%%%%%%%%%%%%%%%%%%%%%%%%%%%%%%%
%\newpage
\vspace{10mm}
\begin{abstract}
We propose a parametrization for two-body nonleptonic $B$ meson
decays, in which the various topologies of amplitudes are counted
in terms of powers of the Wolfenstein parameter $\lambda\sim
0.22$. The weak phases and the amplitudes are determined by
comparing this parametrization with available measurements. It is
possible to obtain the phase $\phi_3$ from the $B\to K\pi$
data up to theoretical uncertainty of $O(\lambda^2)\sim 5\%$. The
recently measured $B_d^0\to\pi^0\pi^0$ branching ratio implies a
large color-suppressed or penguin amplitude, and that the extraction
of the phase $\phi_2$ from the $B\to\pi\pi$ data may suffer
theoretical uncertainty more than the expected one,
$O(\lambda^2)\sim 5\%$.
\end{abstract}
\thispagestyle{empty}
\end{titlepage}

%\newpage

One of the major missions in $B$ physics is to determine the weak
phases in the Kobayashi-Maskawa ansatz for CP violation
\cite{KoMa} . The phase $\phi_1$ can be extracted from the CP
asymmetry in the $B\to J/\psi K_S$ decays in an almost
model-independent way, which arises from the $B$-$\bar B$ mixing.
The application of the isospin symmetry to the $B\to\pi\pi$ decays
\cite{GL} and to the $B\to\rho\pi$ decays \cite{SQ} has been
considered as giving a model-independent determination of the phase
$\phi_2$. However, this strategy in fact suffers the theoretical
uncertainty from the electroweak penguin, which is expected to be
about 5-10\% . The phase $\phi_3$ can be extracted in a
theoretically clean way from the modes involving only tree amplitudes,
such as $B\to \pi D$ \cite{DAD} and $B\to K D$ \cite{gw,ID}.
The difficulty is that one of the modes, such as
$B_d^0\to \pi^- D^+$ or $B^+\to K^+ D^0$, has a very small
branching ratio and is not experimentally feasible \cite{ADS}.
The alternative modes $B\to K^{\ast}D$ \cite{dun} and $B_c\to D_sD$
\cite{FW} improve the feasibility only a bit. It has been pointed out
that the $B^\pm\to K^\pm (D^0\to f)$ and
$B^\pm\to K^\pm ({\bar D}^0\to f)$ amplitudes, with
${\bar D}^0\to f$ being a doubly-Cabibbo suppressed decay, exhibits
a strong interference \cite{ADS,AS,F03}. For this strategy, the
strong phase difference between $D^0\to f$ and $\bar D^0\to f$
is a necessary input. Another possibility is to measure the
$B\to D^*V$ decays for the vector meson $V=\rho$, $K$, $\cdots$,
since an angular analysis involves many observables, which are
sufficient for extracting $\phi_3$ model-independently \cite{LSS}.
%The last two methods, requiring a huge number of $B$ mesons,
%have not yet been realized.

%triangle relations for the $B^\pm_u\to K^\pm
%\{D^0,{\bar D}^0,D^0_+\}$ amplitudes \cite{gw,F03} , where
%$D^0_+$ denotes the CP-even
%eigenstate of the neutral $D$-meson system. This strategy, due to
%the squashed triangles, is experimentally difficult \cite{ADS}.
%Similarly, it is difficult to have a model-independent extraction
%of the angle $\phi_3$ from the $B\to K\pi$, $\pi\pi$ modes
%\cite{GRL,FM,NR,BF}.

Instead of resorting to theoretically clean modes, which are
usually experimentally difficult, one considers the modes with
higher feasibility and tries to constrain the decay amplitudes
and the weak phases. The problem is that available
measurements are usually insufficient to make the constraint,
and theoretical inputs are unavoidable. For example, one adopts
the (imaginary) tree-over-penguin ratio obtained from the
perturbative QCD (PQCD) formalism \cite{LY1,CL,YL,KLS,LUY} or from
the QCD-improved factorization (QCDF) \cite{BBNS}, so that the phase
$\phi_2$ can be extracted from the CP asymmetries of the
$B_d^0\to\pi^+\pi^-$ decays. One may also employ symmetries to relate
the amplitudes of the relevant modes, such as
$SU(3)$ \cite{GRL} and $U$-spin \cite{FGR}, in order to reduce the
number of free parameters. However, the theoretical calculations are
subject to subleading corrections, and the symmetry relations are
broken with unknown symmetry breaking effects. For these strategies
to work, the theoretical uncertainty must be under control.

In this paper we shall propose counting rules for the various
topologies of amplitudes \cite{CC} in two-body nonleptonic $B$ meson
decays in terms of powers of the Wolfenstein parameter
$\lambda\sim 0.22$ \cite{GHL}. The relative importance
among the topological amplitudes has been known from some
physical principles: helicity suppression (color transparency)
implies that tree annihilation (nonfactorizable) contributions
are smaller than leading factorizable emission contributions. Here
we shall assign an explicit power of $\lambda$ to each topology,
such that the relative importance becomes quantitative. This assignment
is supported by the known QCD theories \cite{KLS,BBNS,KL,CKL}, and
differs from that assumed in \cite{GHL}.
We drop the topologies with higher powers of $\lambda$ until the
number of free parameters are equal to the number of available
measurements. The weak phases and the decay amplitudes can then be
solved by comparing the resultant parametrization with experimental
data. Afterwards, it should be examined whether the solved amplitudes
obey the power counting rules. If they do, the extracted weak phases
suffer only the theoretical uncertainty from the neglected
topologies. If not, the inconsistency could be regarded as a warning
to QCD theories for two-body nonleptonic $B$ meson decays. For example,
the long-distance rescattering effect has been neglected in PQCD and
in QCDF. If this effect is important, the hierarchy among the various
topological amplitudes will be destroyed \cite{NGW}. The comparison of
our parametrization with data can tell whether the above assumption is
reliable \cite{CL00}.

As shown below, dropping the electroweak penguin amplitude,
the phase $\phi_2$ can be extracted from
the $B\to \pi\pi$ data. In principle, the theoretical uncertainly of
the ignored amplitudes is around $O(\lambda^2)\sim 5\%$, the same as
in the extraction based on the isospin symmetry \cite{GL}. Similarly,
the phase $\phi_3$ can be best determined from the $B\to K\pi$
data up to the uncertainty from the neglect of the
$O(\lambda^3)\sim 1\%$ tree annihilation and color-suppressed
electroweak amplitudes. Note that the
determination of the phase $\phi_1$ from the $B\to J/\psi K^{(*)}$
decays also bears about 1\% theoretical uncertainty. Certainly, a
CP asymmetry is an $O(\lambda)$ quantity itself. Precisely speaking,
the above determination of $\phi_2$ and $\phi_3$, involving the data of
CP asymmetries, in fact carries the uncertainly of $O(\lambda)\sim 20\%$
and $O(\lambda^2)\sim 5\%$, respectively. Because the
$B\to \pi\pi$, $K\pi$ measurements are not yet complete, we shall drop
more topologies in order to match the currently available data. In this
simple demonstration, we observe that the amplitudes solved from
the $B\to K\pi$ data more or less obey the hierarchy in $\lambda$. That
is, an almost model-independent determination of $\phi_3$ is promising.
The solution from the $B\to\pi\pi$ analysis is, unfortunately, not
consistent with the power counting rules, indicating that the extraction
of $\phi_2$ may suffer theoretical uncertainty larger than stated above.
Hence, our work casts a doubt to the strategy based on the isospin
symmetry \cite{GL} and gives a warning to the
QCD calculations of the $B\to\pi\pi$ modes \cite{LUY,BBNS,BN}.

We start with the $B\to K\pi$ decays. The branching ratio
of a two-body nonleptonic $B$ meson decay is written as
\begin{eqnarray}
B(B\to M_1M_2)=\frac{\tau_B}{16\pi m_B}|A(B\to M_1M_2)|^2\;,
\end{eqnarray}
where the light-meson masses $m_\pi$ and $m_K$ have been
neglected, and the $B$ meson mass and the $B$ meson lifetimes
take the values $m_B = 5.28$
GeV, $\tau_{B^\pm}=1.674\times 10^{-12}$ s, $\tau_{B^0}=1.542\times
10^{-12}$ s. The effective Hamiltonian for the
flavor-changing $b\to s$ transition is \cite{REVIEW},
\begin{equation}
H_{\rm eff}={G_F\over\sqrt{2}}
\sum_{q=u,c}V^*_{qs}V_{qb}\left[C_1(\mu)O_1^{(q)}(\mu)
+C_2(\mu)O_2^{(q)}(\mu)+
\sum_{i=3}^{10}C_i(\mu)O_i(\mu)\right]\;,
\label{hbk}
\end{equation}
with the Cabibbo-Kobayashi-Maskawa (CKM) matrix elements $V$
and the operators,
\begin{eqnarray}
& &O_1^{(q)} = (\bar{s}_iq_j)_{V-A}(\bar{q}_jb_i)_{V-A}\;,\;\;\;\;\;\;\;\;
O_2^{(q)} = (\bar{s}_iq_i)_{V-A}(\bar{q}_jb_j)_{V-A}\;,
\nonumber \\
& &O_3 =(\bar{s}_ib_i)_{V-A}\sum_{q}(\bar{q}_jq_j)_{V-A}\;,\;\;\;\;
O_4 =(\bar{s}_ib_j)_{V-A}\sum_{q}(\bar{q}_jq_i)_{V-A}\;,
\nonumber \\
& &O_5 =(\bar{s}_ib_i)_{V-A}\sum_{q}(\bar{q}_jq_j)_{V+A}\;,\;\;\;\;
O_6 =(\bar{s}_ib_j)_{V-A}\sum_{q}(\bar{q}_jq_i)_{V+A}\;,
\nonumber \\
& &O_7 =\frac{3}{2}(\bar{s}_ib_i)_{V-A}\sum_{q}e_q(\bar{q}_jq_j)_{V+A}\;,
\;\;
O_8 =\frac{3}{2}(\bar{s}_ib_j)_{V-A}\sum_{q}e_q(\bar{q}_jq_i)_{V+A}\;,
\nonumber \\
& &O_9 =\frac{3}{2}(\bar{s}_ib_i)_{V-A}\sum_{q}e_q(\bar{q}_jq_j)_{V-A}\;,
\;\;
O_{10} =\frac{3}{2}(\bar{s}_ib_j)_{V-A}\sum_{q}e_q(\bar{q}_jq_i)_{V-A}\;,
\end{eqnarray}
$i, \ j$ being the color indices.
For the characteristic scale $\mu \sim\sqrt{m_b\bar\Lambda}\sim 1.5$ GeV
involved in two-body $B$ meson decays \cite{KLS},
$\bar\Lambda=m_B-m_b$ being the $B$ meson and $b$ quark mass
difference, the values of the Wilson coefficients are
\begin{eqnarray}
& &C_1 = -0.510\;, \hspace{15mm} C_2 = 1.268\;, \nonumber \\
& &C_3 = 2.7 \times 10^{-2}\;, \hspace{10mm}
C_4 = -5.0 \times 10^{-2}\;, \hspace{10mm}  \nonumber \\
& &C_5 = 1.3 \times 10^{-2}\;, \hspace{10mm}
C_6 = -7.4 \times 10^{-2}\;, \nonumber \\
& &C_7 = 2.6 \times 10^{-4}\;, \hspace{10mm}
C_8 = 6.6 \times 10^{-4}\;, \hspace{10mm} \nonumber \\
& &C_9 = -1.0 \times 10^{-2}\;, \hspace{7mm} C_{10} =
4.0 \times 10^{-3}\;.
\label{wil}
\end{eqnarray}
The above characteristic scale has been confirmed by the dynamical
penguin enhancement exhibited in the $B\to VP$ data \cite{CKL}.
The Wolfenstein parametrization for the CKM matrix is given by
\begin{eqnarray}
\left(\matrix{V_{ud} & V_{us} & V_{ub} \cr
              V_{cd} & V_{cs} & V_{cb} \cr
              V_{td} & V_{ts} & V_{tb} \cr}\right)
&=&\left(\matrix{ 1 - { \lambda^2 \over 2 } & \lambda &
A \lambda^3(\rho - i \eta)\cr
- \lambda & 1 - { \lambda^2 \over 2 } & A \lambda^2\cr
A \lambda^3(1-\rho-i\eta) & -A \lambda^2 & 1 \cr}\right)
%\nonumber\\
=\left(\matrix{ O(1) & O(\lambda) & O(\lambda^4)\cr
O(\lambda) & O(1) & O(\lambda^2)\cr
O(\lambda^3) & O(\lambda^2) & O(1) \cr}\right)\;,
\end{eqnarray}
with the parameters $\lambda = 0.2196 \pm 0.0023$,
$A = 0.819 \pm 0.035$, and $R_b \equiv\sqrt{{\rho}^2  + {\eta}^2}
= 0.41 \pm 0.07$ \cite{LEP}. Note that the product $AR_b\sim 0.3$
should be regarded as being of $O(\lambda)$, and that $|V_{ub}|$
is in fact $O(\lambda^4)$.
The phases $\phi_1$ and $\phi_3$ are defined via
$V_{td}=|V_{td}|\exp(-i\phi_1)$ and $V_{ub}=|V_{ub}|\exp(-i\phi_3)$,
respectively.

Considering all possible topologies of amplitudes,
the $B\to K\pi$ decay amplitudes are given by
\begin{eqnarray}
A(B^+\to K^0\pi^+)&=&P\left(1-\frac{P_{ew}^c}{P}
+\frac{T^a}{P}e^{i\phi_3}\right)\;,
\label{ap1}\\
A(B_d^0\to K^+\pi^-)&=&-P\left(1-\frac{P_{ew}^a}{P}
+\frac{T}{P}e^{i\phi_3}\right)\;,
\label{bp1}\\
\sqrt{2}A(B^+\to K^+\pi^0)&=&-P\left[1+\frac{P_{ew}}{P}
+\left(\frac{T}{P}+\frac{C}{P}+\frac{T^a}{P}\right)e^{i\phi_3}\right]\;,
\label{app1}\\
\sqrt{2}A(B_d^0\to K^0\pi^0)&=&P\left(1
-\frac{P_{ew}}{P}-\frac{P_{ew}^c}{P}-\frac{P_{ew}^a}{P}
-\frac{C}{P}e^{i\phi_3}\right)\;, \label{bpp1}
\end{eqnarray}
which satisfy the quadrangle relation,
\begin{eqnarray}
A(B^+\to K^0\pi^+)+\sqrt{2}A(B^+\to K^+\pi^0)= A(B_d^0\to
K^+\pi^-)+\sqrt{2}A(B_d^0\to K^0\pi^0)\;.
\end{eqnarray}
The amplitude $P_{ew}$ ($P_{ew}^c$, $P_{ew}^a$) comes from the
color-allowed (color-suppressed, annihilation) topology through
the electroweak penguin operators. The amplitude $P$ includes the
emission and annihilation topologies through both the QCD and
electroweak penguins:
\begin{eqnarray}
P=P_{QCD}+e_uP_{ew}^c+e_uP_{ew}^a\;,
\label{p}
\end{eqnarray}
with the $u$ quark charge $e_u=2/3$.
The amplitude $T$ ($C$, $T^a$) comes from the color-allowed
(color-suppressed, annihilation) topology through the tree
operators.
The penguin contributions from the $c$ quark loop can be included
using the relation $V^*_{cs}V_{cb}=-V^*_{us}V_{ub}
-V^*_{ts}V_{tb}$, and the expressions in
Eqs.~(\ref{Map1})-(\ref{Mbpp1}) remain unchanged.

It has been shown in PQCD that a nonfactorizable amplitude
$M^{nf}$, a factorizable annihilation amplitude $F^a_{(V-A)}$
from the $(V-A)(V-A)$ current, and a factorizable annihilation
amplitude $F^a_{(V+A)}$ from the $(V-A)(V+A)$ current
are suppressed, compared to the leading factorizable emission
amplitude $F^e$, by the factors of\cite{LU},
\begin{eqnarray}
\frac{M^{nf}}{F^e}\sim \left[\ln\frac{m_B}{\Lambda_{\rm
QCD}}\right]^{-1}\sim \lambda\;,\;\;\;\;
\frac{F^a_{(V-A)}}{F^e}\sim \frac{\Lambda_{\rm QCD}}{m_B}\sim
\lambda^2\;,\;\;\;\;
\frac{F^a_{(V+A)}}{F^e}\sim \frac{2m_0}{m_B}\sim \lambda^0\;,
\label{pcr0}
\end{eqnarray}
respectively, where $m_0$ is the chiral enhancement scale, and
the CKM matrix elements and the Wilson
coefficients are excluded. We list the power counting rules
for the Wilson coefficients in Eq.~(\ref{wil}),
\begin{eqnarray}
&&O(1): a_1\;, \nonumber  \\
&&O(\lambda): a_2\;, 1/N_c \;, \nonumber\\
&&O(\lambda^2): C_4\;, C_6\;, a_4\;, a_6 \;, \nonumber \\
&&O(\lambda^3): C_3\;, C_5\;, C_9\;, a_3\;, a_5\;, a_9\;,\nonumber\\
&&O(\lambda^4): C_{10}\;, \nonumber\\
&&O(\lambda^5): C_7\;, C_8\;, a_7\;, a_8\;,a_{10} \;,\label{pcr}
\end{eqnarray}

with $a_1=C_2+C_1/N_c$, $a_2=C_1+C_2/N_c$,
$a_i=C_i+C_{i+1}/N_c$ for $i=3$, 5, 7, 9, and
$a_i=C_i+C_{i-1}/N_c$ for $i=4$, 6, 8, 10.

According to Eqs.~(\ref{pcr0}) and (\ref{pcr}),
we assign the powers of $\lambda$ to the following ratios of the
various topological amplitudes:
\begin{eqnarray}
& &\frac{T}{P}\sim \frac{V_{us}V^*_{ub}}{V_{ts}V^*_{tb}}
\frac{a_1}{a_{4,6}}
\sim \lambda\;,\;\;\;\;
\frac{P_{ew}}{P}\sim\frac{a_9}{a_{4,6}}\sim \lambda\;,\;\;\;\;
\frac{C}{T}\sim \frac{a_2}{a_1}\sim\lambda\;,
\nonumber\\
& &\frac{T^a}{T}\sim \frac{F^a_{(V-A)}}{F^e}\sim
\frac{M^{nf}}{F^e}\frac{C_1}{a_1N_c}\sim
\lambda^2\;,
\nonumber\\
& &\frac{P_{ew}^c}{P}\sim\frac{a_{8,10}}{a_{4,6}}
\sim\frac{M^{nf}}{F^e}\frac{C_9}{a_{4,6}N_c}\sim \lambda^3\;,
\nonumber\\
& &\frac{P_{ew}^a}{P}\sim\frac{F^a_{(V+A)}}{F_e}
\frac{a_{8,10}}{a_{4,6}}
\sim\frac{M^{nf}}{F^e}\frac{C_9}{C_{4,6}N_c}\sim \lambda^3\;.
\label{pow}
\end{eqnarray}
For the latter three ratios, we present the power counting
rules derived from both the factorizable and nonfactorizable
contributions, which are of the same order of magnitude.
Compared to the power counting rules in \cite{GHL} based on the
conventional scale $\mu\sim m_b$, $P_{ew}^c/P$ is down by one more
power of $\lambda$ due to $a_{10}\sim O(\lambda^5)$ in PQCD.

Whether a factorizable amplitude or a nonfactorizable amplitude is
important depends on the decay modes. In the $B\to K\pi$ case,
$C$ mainly comes from the factorizable color-suppressed diagrams,
since there is a strong cancellation between a pair of
nonfactorizable diagrams. The factorizable and nonfactorizable
annihilation contributions to $T^a$, $P_{ew}^c$, and $P_{ew}^a$
are of the same order of magnitude as shown in Eq.~(\ref{pow}).
In the $B\to D\pi$ decays,
$C$, being of the same order of magnitude as $T$, mainly comes from
the nonfactorizable color-suppressed
diagrams, since the above cancellation does not exist \cite{YL,KKL}.
For $T^a$ in the $B\to D\pi$ case, the
nonfactorizable diagrams dominate, because of
\begin{eqnarray}
\frac{M^{nf}}{F^e}\frac{C_2}{a_1N_c}\sim \lambda^2 \gg
\frac{F^a_{(V-A)}}{F^e}\frac{a_2}{a_1}\sim \lambda^3\;.
\label{poc}
\end{eqnarray}

Employing the reparametrizations,
\begin{eqnarray}
P-P_{ew}^a\to P\;,\;\;\;
P_{ew}+P_{ew}^a\to P_{ew}\;,\;\;\;
P_{ew}^c-P_{ew}^a\to P_{ew}^c\;,
\end{eqnarray}
we arrive at the most general parametrization of the
$B\to K\pi$ decay amplitudes,
\begin{eqnarray}
A(B^+\to K^0\pi^+)&=&P\left(1-\frac{P_{ew}^c}{P}
+\frac{T^a}{P}e^{i\phi_3}\right)\;,
\label{Map1}\\
A(B_d^0\to K^+\pi^-)&=&-P\left(1+\frac{T}{P}e^{i\phi_3}\right)\;,
\label{Mbp1}\\
\sqrt{2}A(B^+\to K^+\pi^0)&=&-P\left[1+\frac{P_{ew}}{P}
+\left(\frac{T}{P}+\frac{C}{P}
+\frac{T^a}{P}\right)e^{i\phi_3}\right]\;,
\label{Mapp1}\\
\sqrt{2}A(B_d^0\to K^0\pi^0)&=&P\left(1
-\frac{P_{ew}}{P}-\frac{P_{ew}^c}{P}
-\frac{C}{P}e^{i\phi_3}\right)\;. \label{Mbpp1}
\end{eqnarray}
There are totally 6 independent amplitudes, namely, 11 unknowns, because
an overall phase can always be removed. Hence, we choose the amplitude
$P$ as a positive real value. Plus the weak phase $\phi_3$, the
12 unknowns are definitely more than the 9 experimental inputs:
the branching ratios and the direct CP asymmetries of the four
modes, and the mixing-induced CP asymmetry of the $B_d^0\to K^0\pi^0$
mode. Dropping the $O(\lambda^3)$ terms, $T^a/P$ and $P_{ew}^c/P$,
we have 8 unknowns. Then the data of the direct CP asymmetry in
the $B^+\to K^0\pi^\pm$ decays should be excluded for consistency.
Hence, we have 8 experimental inputs, and thus all unknowns can be
solved exactly assuming the phase $\phi_1$ is already known
from the measurement of the mixing-induced CP asymmetry in the
$B\to J/\psi K^{(*)}$ modes. The determination of $\phi_3$
from this parametrization is then accurate up to the
theoretical uncertainty of $O(\lambda^2)\sim 5\%$.

We emphasize the consequence from the different power counting rules
in \cite{GHL} and in this work: the smaller $P_{ew}^c$ is crucial for
claiming that the determination
of $\phi_3$ from the $B\to K \pi$ data is accurate up to 5\%
theoretical uncertainty. Following the counting rules in \cite{GHL},
both $C$ and $P_{ew}^c$ will be included at $O(\lambda^2)$, such that
the 10 unknowns are more than the 9 available measurements. In this
case we can not solve for $C$ and $P_{ew}^c$ exactly, and have to
rely on symmetry relations to reduce the number of unknowns. It is
then difficult to estimate the involved theoretical uncertainty. Using
the counting rules in Eq.~(\ref{pow}), which are supported by the PQCD
calculation, we include only $C$ at $O(\lambda^2)$, and the number of
unknowns can be equal to the number of measurements. Solving for $C$,
and assuring that the solution obeys our counting rule as a
self-consistency check, the uncertainty from the neglected topologies
is under control.

The measurement of the time-dependent asymmetry in the
$B_d^0\to K_S\pi^0$ decay still suffers a large error. To demonstrate
our method, we reduce the number of unknowns by further
dropping the $O(\lambda^2)$ terms, $C/P$, arriving at
\begin{eqnarray}
A(B^+\to K^0\pi^+)&=&P\;,
\label{Map2}\\
A(B_d^0\to K^+\pi^-)&=&-P\left(1+\frac{|T|}{P}e^{i\phi_3}
e^{i\delta_T}\right)\;,
\label{Mbp2}\\
\sqrt{2}A(B^+\to K^+\pi^0)&=&-P\left(1+\frac{|P_{ew}|}{P}
e^{i\delta_{ew}}
+\frac{|T|}{P}e^{i\phi_3}e^{i\delta_T}\right)\;,
\label{Mapp2}\\
\sqrt{2}A(B_d^0\to K^0\pi^0)&=&P\left(1
-\frac{|P_{ew}|}{P}e^{i\delta_{ew}}\right)\;,
\label{Mbpp2}
\end{eqnarray}
where $\delta_T$ and $\delta_{ew}$ denote the strong phases of
$T$ and $P_{ew}$, respectively. The $B\to K\pi$ decay amplitudes
in Eqs.~(\ref{Map2})-(\ref{Mbpp2}) are the expansion up to the
power of $\lambda$, at which the determination of $\phi_3$
suffers the theoretical uncertainty of $O(\lambda)\sim 20\%$.

We shall solve for the 6 unknowns: $P$, $|P_{ew}|$, $|T|$,
$\phi_3$, and the strong phases $\delta_{ew}$ and $\delta_T$,
from the 6 experimental inputs \cite{KS,BABAR},
\begin{eqnarray}
& &{\rm Br}(B^\pm\to K^0\pi^\pm)=(20.6\pm 1.4)\times 10^{-6}\;,
%=(18.1\pm 1.7)\times 10^{-6}\;,
\nonumber\\
& &{\rm Br}(B_d^0\to K^\pm\pi^\mp) =(18.2\pm 0.8)\times 10^{-6}\;,
%=(18.5\pm 1.0)\times 10^{-6} \;,
\nonumber\\
& &{\rm Br}(B^\pm\to K^\pm\pi^0)=(12.8\pm 1.1 \times 10^{-6}\;,
%=(12.7\pm 1.2\times 10^{-6}\;,
\nonumber\\
& &{\rm Br}(B_d^0\to K^0\pi^0)=(11.5\pm 1.7) \times 10^{-6} \;,
%=(10.2\pm 1.5)\times 10^{-6} \;,
\nonumber\\
& &{\cal A}(B_d^0\to K^\pm\pi^\pm)=-(10.2\pm 5.0)\%\;,
\nonumber\\
& &{\cal A}(B^\pm\to K^\pm\pi^0)=-(9.0\pm 9.0)\%\;.
\label{inp}
\end{eqnarray}
The $B^\pm\to K^0\pi^\pm$ and $B_d^0\to K^0\pi^0$ modes indeed
have very small direct CP asymmetries, consistent with the
parametrization in Eqs.~(\ref{Map2})-(\ref{Mbpp2}). The bounds on
the various amplitudes and phases can be derived unambiguously
from Eq.~(\ref{inp}).

%%%%%%%%%%%%%%%%%%%%%%%%%%%%%%%%%%%%%%%%%
\begin{figure}[t]
\begin{center}
\epsfig{file=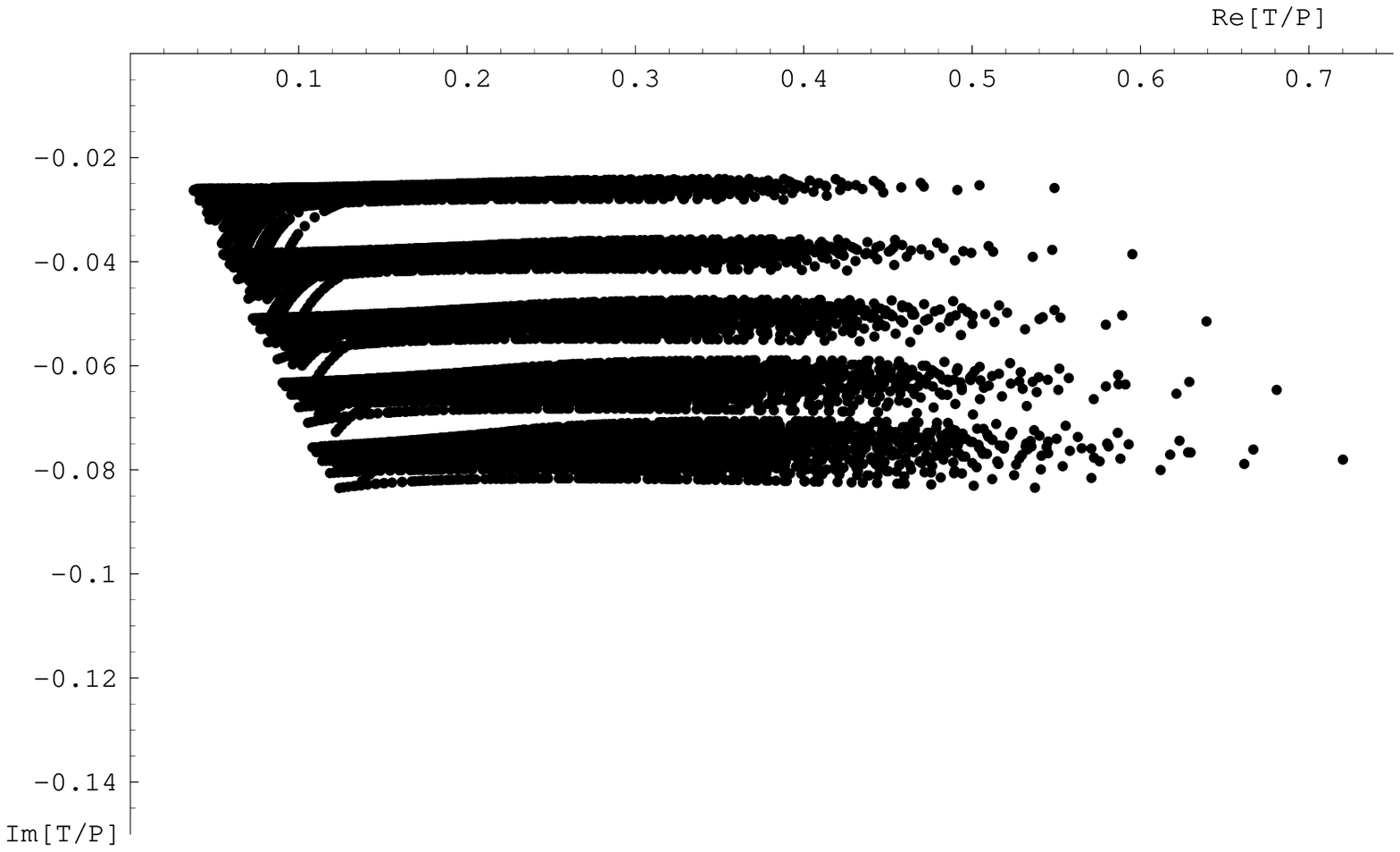,width=9cm}
\end{center}
\caption{The allowed range of $T/P$ determined from the $B\to
K\pi$ data.} \label{tp}
\end{figure}
%%%%%%%%%%%%%%%%%%%%%%%%%%%%%%%%%%%%%%%%%%%%

%%%%%%%%%%%%%%%%%%%%%%%%%%%%%%%%%%%%%%%%%
\begin{figure}[t]
\begin{center}
\epsfig{file=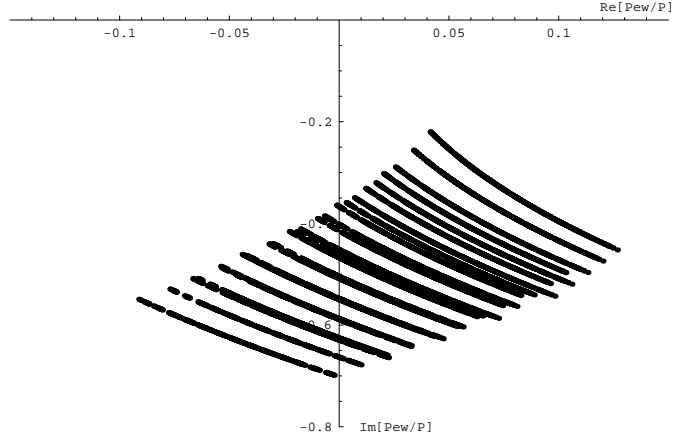,width=9cm}
\end{center}
\caption{The allowed range of $P_{ew}/P$ determined from the $B\to
K\pi$ data.} \label{pew}
\end{figure}
%%%%%%%%%%%%%%%%%%%%%%%%%%%%%%%%%%%%%%%%%%%%

The allowed ranges of the ratios $T/P$ and $P_{ew}/P$
are exhibited in Figs.~\ref{tp} and \ref{pew}, respectively.
The prescription for deriving the two figures is briefly explained
below. The data for each branching ratio and for each CP asymmetry are
expressed as a set, whose elements are the central value with
$+1\times$ error bar, $0\times$ error bar, and $-1\times$ error bar.
For a combination of the element
from each set, we solve the coupled equations, and the solution
is represented by a dot in the figure. Scanning all the
combinations, we obtain the ranges in the figures. The central
values of the solutions are
\begin{eqnarray}
& &\frac{|T|}{P}=0.23\;,\;\;\;\delta_T=-13^o\;,
\nonumber\\
& &\frac{|P_{ew}|}{P}=0.50\;,\;\;\;\delta_{ew}=-88^o\;.
\end{eqnarray}
The above result of $T/P$ is in agreement with the PQCD
prediction, $T/P\sim 0.20\exp(- 27^oi)$ \cite{KLS,KS,US,KS03},
while the central values of $|P_{ew}|/P$ and of $\delta_{ew}$
differ from the PQCD prediction, $|P_{ew}|/P\sim 0.2$ and
$\delta_{ew}\approx \delta_T$, respectively. The latter PQCD
prediction is consistent with the almost model-independent
relation between the electroweak penguin and tree amplitudes
obtained in \cite{NGW,NR98}. The ratio $|P_{ew}|/P=0.5$ and the 
nearly $90^o$ phase between $P_{ew}$ and $P$ in the above fit
have been speculated in \cite{BN,GR03,Y03}. We also derive the allowed
ranges $0.06 <|T|/P<0.72$ and $0.22<|P_{ew}|/P<0.70$, implying
that the extracted ratios $|T|/P$ and $|P_{ew}|/P$ deviate a bit
from the power counting rules in Eq.~(\ref{pow}). Hence, the $B\to
K\pi$ data are indeed puzzling, especially from the viewpoint
of the dramatically different strong phases $\delta_{ew}$ and 
$\delta_T$ shown in Figs.~\ref{tp} and \ref{pew}. Because of the 
large central values of $|P_{ew}|/P$ and of $\delta_{ew}$, a strong 
hint of new physics has been claimed in \cite{BN,BFR,Y03}. 
A more convincing examination of the self-consistency can be made
by solving for the amplitude $C$, when more complete data are
available. At last, the central value and the allowed range of the
phase $\phi_3$ are given by
\begin{eqnarray}
\phi_3=102^o\;,\;\;\; 26^o<\phi_3<151^o\;,
\end{eqnarray}
respectively, with the theoretical uncertainty of about 20\%.

%However, viewing
%that $|P_{ew}|/P$ could be as low as 0.22 and $\delta_{ew}$ 
%could be equal to $\delta_T$ within $1\sigma$, we would like
%to be more prudential, and suggest to wait for more precise data.

We emphasize that our fitting differs from the global fitting
based on the QCDF approach \cite{BBNS,DSY}. For example, the penguin
contributions have been split into the factorizable type depending
on a transition form factor, the nonfactorizable type depending on
the imaginary infrared cutoff $\rho_H$ for an end-point singularity,
and the annihilation type depending on the imaginary infrared cutoff
$\rho_A$ in QCDF. Taking into account only the $B\to PP$ modes, such as
$B\to K\pi$ and $\pi\pi$, the fitting result of the phase
$\phi_3\sim 110^o$ \cite{Be02} is close to that extracted in this work.
Our method also differs from those based on the isospin relations
\cite{iso}, with which some combinations of the
$B\to K\pi$ branching ratios can be described by the
functions of the parameters $P_{ew}/T$, $T/P$ and the
relative strong phases. The $SU(3)$ flavor symmetry is then
employed to fix $P_{ew}/T$ and $T/P$. Finally, only the strong
phases and the weak phase $\phi_3$ are treated as unknowns, and
determined by the data. The conclusion is similar: the $B\to
K\pi$ data favor $\phi_3\ge 90^o$. Our approach does not
rely on the $SU(3)$ symmetry, and the ratios
$P_{ew}/T$ and $T/P$ are treated as unknowns.
Including the $B\to VP$ modes in the QCDF fitting,
the value of $\phi_3$ could be smaller than $90^o$ \cite{DSY}.
Using the parametrization for the the $B\to VP$ modes based on
SU(3) flavor symmetry, an phase $\phi_3< 90^o$ was also obtained
\cite{CRG}. In a forthcoming paper we shall apply our parametrization
to the $B\to VP$ modes, and make a comparison with the above works.
%from the global fitting based on the naive
%factorization assumption \cite{WS}.

Next we apply our method to the $B\to\pi\pi$ decays.
Considering all possible topologies of amplitudes, their decay
amplitudes are given by
\begin{eqnarray}
\sqrt{2}A(B^+\to \pi^+\pi^0)&=&-T\left[1+\frac{C}{T}
+\left(\frac{P_{ew}}{T}+\frac{P_{ew}^c}{T}
+\frac{P_{ew}^a}{T}\right)e^{i\phi_2}\right]\;,
\label{a1}\\
A(B_d^0\to \pi^+\pi^-)&=&-T\left(1+\frac{T^a}{T}
+\frac{P}{T}e^{i\phi_2}\right)\;,
\label{b1}\\
\sqrt{2}A(B_d^0\to \pi^0\pi^0)&=&T\left[\left(
\frac{P}{T}-\frac{P_{ew}}{T}-\frac{P_{ew}^c}{T}
-\frac{P_{ew}^a}{T}\right)
e^{i\phi_2}-\frac{C}{T}
+\frac{T^a}{T}\right]\;,
\label{bpi1}
\end{eqnarray}
which satisfy the triangle relation,
\begin{eqnarray}
\sqrt{2}A(B^+\to \pi^+\pi^0)=A(B_d^0\to
\pi^+\pi^-)+\sqrt{2}A(B_d^0\to \pi^0\pi^0)\;.
\end{eqnarray}
In the above expressions the amplitude $P$ has been defined in
Eq.~(\ref{p}), and the annihilation contribution $T^a$ comes
only from the nonfactorizable diagrams. Based on Eqs.~(\ref{pcr0}),
(\ref{pcr}) and (\ref{pow}), we assign the power counting rules to
the following ratios of the topological amplitudes:
\begin{eqnarray}
& &\frac{P}{T}\sim \frac{V_{td}V^*_{tb}}
{V_{ud}V^*_{ub}}\frac{a_{4,6}}{a_1}\sim\lambda\;,\;\;\;\;
\frac{C}{T}\sim \lambda\;,
\nonumber\\
& &\frac{P_{ew}}{T}\sim \lambda^2\;,\;\;\;\;
\frac{T^a}{T}\sim
\frac{M_{nf}}{M_e}\frac{C_2}{a_1N_c}\sim
\lambda^2\;,
\nonumber\\
& &\frac{P_{ew}^c}{T}\sim \frac{P_{ew}^a}{T}\sim\lambda^4\;.
\label{po}
\end{eqnarray}

Employing the reparametrizations,
\begin{eqnarray}
T+T^a\to T\;,\;\;\; C-T^a\to C\;,\;\;\;
P_{ew}+P_{ew}^c+P_{ew}^a\to P_{ew}\;,
\end{eqnarray}
the most general parametrizations of the $B\to\pi\pi$ decay
amplitudes are written as
\begin{eqnarray}
\sqrt{2}A(B^+\to \pi^+\pi^0)&=&-T\left[1+\frac{C}{T}
+\frac{P_{ew}}{T}e^{i\phi_2}\right]\;,
\label{Ma1}\\
A(B_d^0\to \pi^+\pi^-)&=&-T\left(1
+\frac{P}{T}e^{i\phi_2}\right)\;,
\label{Mb1}\\
\sqrt{2}A(B_d^0\to \pi^0\pi^0)&=&T\left[\left(
\frac{P}{T}-\frac{P_{ew}}{T}\right)
e^{i\phi_2}-\frac{C}{T}\right]\;.
\label{Mbpi1}
\end{eqnarray}
There are 4 independent amplitudes, namely, 8 parameters including
the phase $\phi_2$, which are more than the available measurements.
Neglecting the $O(\lambda^2)$ term, $P_{ew}/T$,
the resultant expressions are the same as in \cite{GHL}:
\begin{eqnarray}
\sqrt{2}A(B^+\to \pi^+\pi^0)&=&-T\left(1+\frac{|C|}{T}
e^{i\delta_C}\right)\;,
\label{Ma11}\\
A(B_d^0\to \pi^+\pi^-)&=&-T\left(1
+\frac{|P|}{T}e^{i\phi_2}e^{i\delta_P}\right)\;,
\label{Mb11}\\
\sqrt{2}A(B_d^0\to \pi^0\pi^0)&=&T\left(
\frac{|P|}{T}e^{i\phi_2}e^{i\delta_P}
-\frac{|C|}{T}e^{i\delta_C}\right)\;,
\label{Mbpi11}
\end{eqnarray}
for which we have 6 unknowns $T$, $|C|$, $|P|$, $\delta_C$,
$\delta_P$ and $\phi_2$. Similarly, we have removed the strong
phase of $T$, and assumed it to be real and positive.

In this case we have to exclude the data of the direct CP
asymmetry in the $B^+\to \pi^+\pi^0$ decay, and 6 experimental
inputs are relevant: the three CP-averaged branching ratios, the
direct and mixing-induced CP asymmetries in $B_d^0\to
\pi^+\pi^-$, and the direct CP asymmetry in $B_d^0\to
\pi^0\pi^0$. At this level of accuracy, our parametrization
is equivalent to that based on the isospin triangle
\cite{GL,LYQ}, in which the electroweak penguin contribution to the
$B^+\to \pi^+\pi^0$ decay is also ignored. We mention that the
electroweak penguin amplitude has been included in the isospin
analysis of the $B\to\pi\pi$ decays, and that the CP asymmetry in
the $B^\pm\to \pi^\pm\pi^0$ modes still vanishes \cite{GPY}.
%It seems that the
%theoretical uncertainty in the determination of $\phi_2$ from the
%$B\to\pi\pi$ decays must be larger than in the determination of
%$\phi_3$ from the $B\to K\pi$ decays.
After extracting $\phi_2$ from the $B\to\pi\pi$ data and $\phi_3$ from
the $B\to K\pi$ data, we can check whether they, together with $\phi_1$
from the $B\to J/\psi K^{(*)}$ data, satisfy the unitarity
constraint, when the data precision improves.

The time-dependent CP asymmetry of the $B_d^0\to
\pi^+\pi^-$ mode is expressed as
\begin{eqnarray}
{\cal A}(B_d^0(t)\to\pi^+\pi^-)&\equiv&
\frac{B({\bar B}^0_d(t)\to\pi^+\pi^-)-
B(B_d^0(t)\to\pi^+\pi^-)}{B({\bar B}^0_d(t)\to\pi^+\pi^-)
+B(B_d^0(t)\to\pi^+\pi^-)}
\nonumber\\
&=&- C_{\pi\pi}\cos(\Delta M_dt)+S_{\pi\pi}\sin(\Delta
M_dt)\;,\label{CPASY}
\end{eqnarray}
where the direct asymmetry $ C_{\pi\pi}$ and the mixing-induced
asymmetry $S_{\pi\pi}$ are defined by
\begin{eqnarray}
C_{\pi\pi}={1-|\lambda_{\pi\pi}|^2 \over
1+|\lambda_{\pi\pi}|^2}\;, \hspace{20mm}
S_{\pi\pi}={2\,Im(\lambda_{\pi\pi}) \over
1+|\lambda_{\pi\pi}|^2}\;,
\end{eqnarray}
respectively, with the factor,
\begin{eqnarray}
\lambda_{\pi\pi}  =e^{2i\phi_2} {1+e^{-i\phi_2} P/T \over
1+e^{i\phi_2}P/T} \;.
\end{eqnarray}
The data are summarized as \cite{HJ},
\begin{eqnarray}
& &{\rm Br}(B^\pm\to \pi^\pm\pi^0) =(5.2\pm 0.8)\times 10^{-6}\;,
\nonumber\\
& &{\rm Br}(B_d^0\to \pi^\pm\pi^\mp) =(4.6\pm 0.4)\times 10^{-6}\;,
\nonumber\\
& &{\rm Br}(B_d^0\to \pi^0\pi^0) =(1.97\pm 0.47)\times 10^{-6}\;,
\nonumber\\
& &C_{\pi\pi}=-(38\pm 16)\%\;,\;\;\;\;
S_{\pi\pi}=-(58\pm 20)\%\;.
\label{pp}
\end{eqnarray}
Since the data of the direct CP asymmetry in the $B_d^0\to \pi^0\pi^0$
mode is not yet available, we shall assign a plausible range to it,
\begin{eqnarray}
{\cal A}(B_d^0\to \pi^0\pi^0) =(-50\sim +50)\%\;.
\end{eqnarray}

%${\cal A}^{\rm dir}(B_d\to\pi^0K_{\rm S})$ and
%${\cal A}^{\rm mix}(B_d\to\pi^0K_{\rm S})$ are due to ``direct'' and
%``mixing-induced'' CP violation, respectively. These two quantities can
%be expressed in terms of the $B_d^0\to K^0\pi^0$ amplitude  \cite{RF}.
%If the tree amplitude was neglected, ${\cal A}^{\rm dir}$ vanishes, and
%\begin{equation}
%{\cal A}^{\rm mix}(B_d\to\pi^0K_{\rm S})=\sin(2\phi_1)=
%{\cal A}^{\rm mix}(B_d\to J/\psi\,K_{\rm S})\;.
%\label{CP-rel}
%\end{equation}
%It turns out that the mixing-induced CP asymmetry is not related
%to Eq.~(\ref{Mbpp1}).

%%%%%%%%%%%%%%%%%%%%%%%%%%%%%%%%%%%%%%%%%
\begin{figure}[t]
\begin{center}
\epsfig{file=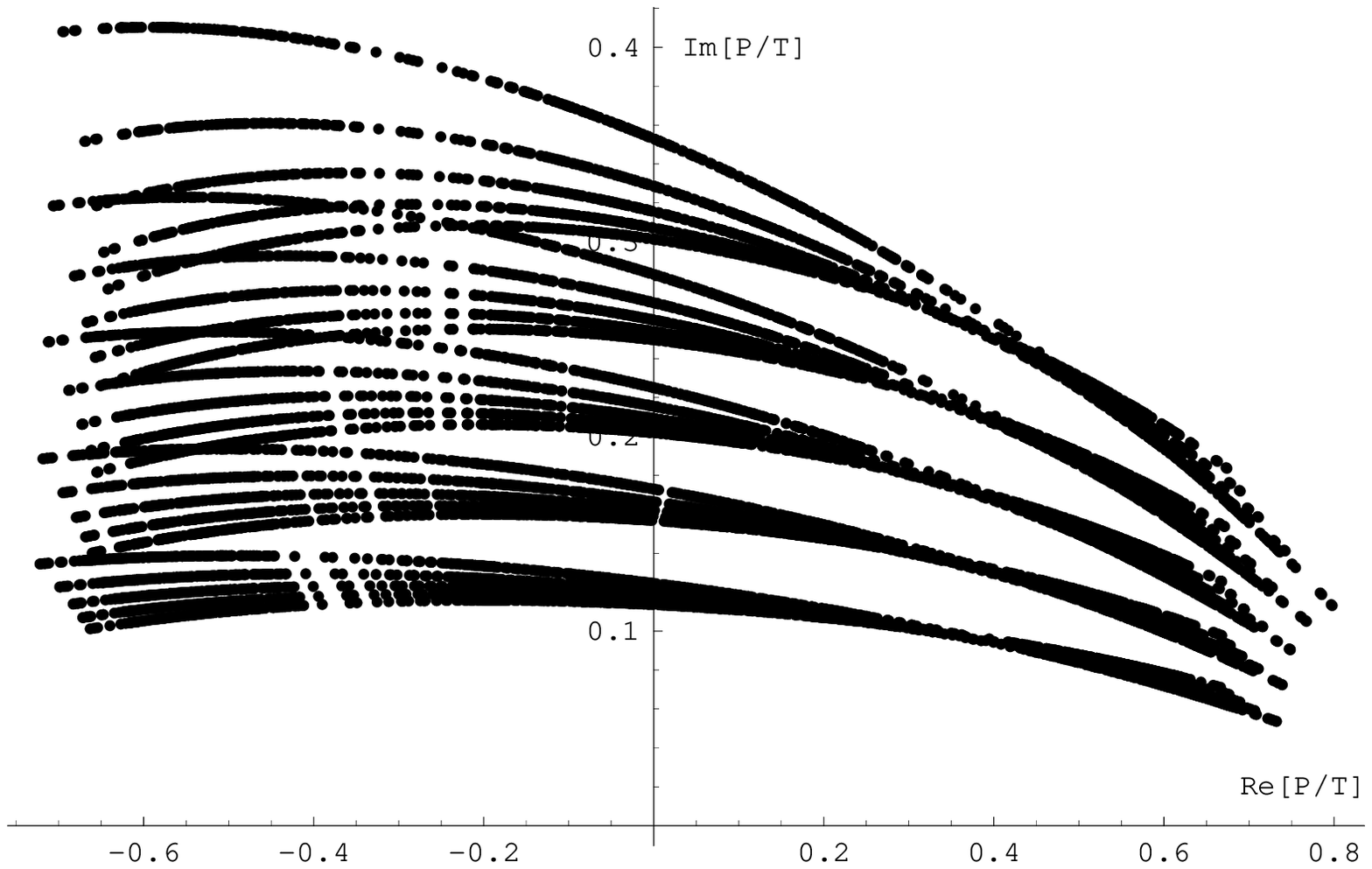,width=9cm}
\end{center}
\caption{The allowed range of $P/T$ determined from the $B\to
\pi\pi$ data.} \label{pip}
\end{figure}
%%%%%%%%%%%%%%%%%%%%%%%%%%%%%%%%%%%%%%%%%%%%

%%%%%%%%%%%%%%%%%%%%%%%%%%%%%%%%%%%%%%%%%
\begin{figure}[t]
\begin{center}
\epsfig{file=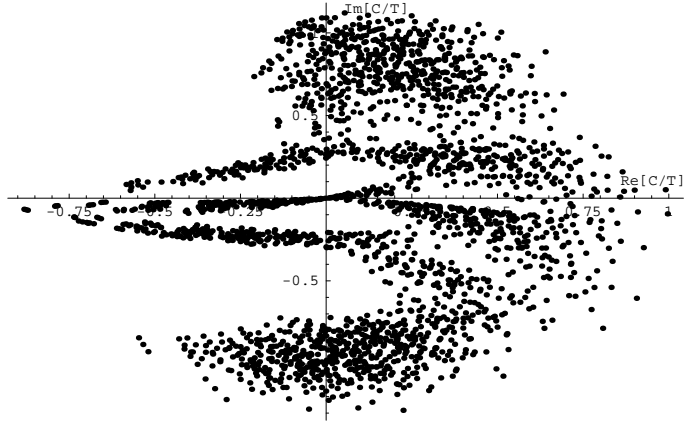,width=9cm}
\end{center}
\caption{The allowed range of $C/T$ determined from the $B\to
\pi\pi$ data.} \label{pic}
\end{figure}
%%%%%%%%%%%%%%%%%%%%%%%%%%%%%%%%%%%%%%%%%%%%

The central values of the measured $B^\pm\to \pi^\pm\pi^0$ and
$B_d^0\to \pi^\pm\pi^\mp$ branching ratios are close to each
other, implying that either $C$ is large and constructive in order
to enhance the $B^\pm\to \pi^\pm\pi^0$ modes, or $P$ is large and
destructive (after including the weak phase $\phi_2$) in order to
suppress the $B_d^0\to \pi^\pm\pi^\mp$ modes \cite{BN}. In either
case the $B_d^0\to \pi^0\pi^0$ branching ratio exceeds the
expected order of magnitude, $O(10^{-7})$. There exist four
solutions associated with each set of data input: two solutions
correspond to the large $C$ and $P$ cases, and the other two are
the reflections of the first two with respect to the $B^\pm\to
\pi^\pm\pi^0$ side of the isospin triangle. Note that the
relations of the phase $\phi_2$ to the measured quantities have
been given in \cite{X} without numerical results. Here we shall
not present the central values of the solutions, because the
central values of the experimental data of the $B_d^0\to
\pi^0\pi^0$ direct CP asymmetry are not yet available.

%from the central values of the data:
%\begin{eqnarray}
%& &\frac{P}{T}=0.21e^{96^oi}\;,\;\;\;
%\frac{C}{T}=0.99e^{-84^oi}\;,\;\;\;\phi_2=107^o\;, \label{s1}\\
%& &\frac{P}{T}=0.21e^{76^oi}\;,\;\;\;
%\frac{C}{T}=0.83e^{76^oi}\;,\;\;\;\phi_2=111^o\;, \label{s2}\\
%& &\frac{P}{T}=0.67e^{162^oi}\;,\;\;\;
%\frac{C}{T}=0.49e^{-18^oi}\;,\;\;\;\phi_2=72^o\;, \label{s3}\\
%& &\frac{P}{T}=0.67e^{10^oi}\;,\;\;\;
%\frac{C}{T}=0.16e^{170^oi}\;,\;\;\;\phi_2=147^o\;. \label{s4}
%\end{eqnarray}
%The solutions in Eqs.~(\ref{s1}) and (\ref{s2})
%[Eqs.~(\ref{s3}) and (\ref{s4})]
%stated above. Moreover, Eq.~(\ref{s1}) is a reflection of Eq.~(\ref{s2})
%Similarly, Eqs.~(\ref{s3}) and (\ref{s4}) are reflections to each other.
%It is then realized why four solutions exist, differing from the
%$B\to K\pi$ case with a single solution.
%Equations (\ref{s1})-(\ref{s4})  This inconsistency is also
%obvious from

The ranges of $P/T$ and $C/T$, shown in Figs.~\ref{pip} and
\ref{pic}, respectively, collect all allowed solutions. These
ranges indicate that the hierarchy in Eq.~(\ref{po}) is not
satisfied, since both $|P|/T$ and $|C|/T$ can be as large as 1,
much greater than $O(\lambda)\sim 0.22$. There is then no reason
for believing that the effect of the electroweak penguin would be
as small as $O(\lambda^2)\sim 5\%$ according to the relation
between $P_{ew}$ and $T$ \cite{NGW,NR98}. Our analysis implies
that the extraction of $\phi_2$ from the $B\to\pi\pi$ data based
on the isospin symmetry may suffer the theoretical uncertainty
more than expected. It also casts a doubt to the PQCD (also QCDF)
calculation of the $B\to\pi\pi$ decays. To complete our numerical
study, we present the allowed range of $\phi_2$ corresponding to
the data in Eq.~(\ref{pp}),
\begin{eqnarray}
51^o < \phi_2 < 176^o\;.
\end{eqnarray}
As explained above, the theoretical uncertainty associated with
the above range may not be under control.

When the data become more precise, and when the data of more CP
asymmetries, such as the mixing-induced CP asymmetry in the
$B_d^0\to K_S\pi^0$ mode, are available, the allowed range will
shrink, and the theoretical uncertainty can reduce. Our
method then tells whether the $B\to K\pi$ data indicate a solid
signal of new physics. Besides,
our parametrization extends straightforwardly to
the other relevant modes, such as $B\to K^*\pi$, from which the
phase $\phi_3$ can also be extracted \cite{Sun}. Considering
the overlap of the extractions from different modes, the
allowed ranges of the decay amplitudes and of $\phi_3$ will shrink
too. An evaluation of the
next-to-leading-order corrections to the $B\to\pi\pi$ decays in the
PQCD framework is now in progress, whose result will
clarify whether the large $|P|$ or $|C|$ is understandable. If not,
new dynamics, such as the rescattering effect, might be important.
The $B\to\pi\pi$ decays and the extraction of the phase $\phi_2$
then demand more theoretical effort.

\vskip 1.0cm
We thank R. Fleischer, A. H\"ocker, Y.Y. Keum, H. Lacker and D. London
for useful discussions.
This work was supported by the National Science Council of
R.O.C. under the Grant No. NSC-92-2112-M-001-030 and by the National
Center for Theoretical Sciences of R.O.C..

\end{document}